\documentclass[]{article}

\usepackage{graphicx}
\usepackage{dcolumn}
\usepackage{bm}

\begin{document}

\title{Implications of two-body fragment decay 
       for the interpretation of emission chronology 
       from velocity-gated correlation functions}

\author{Johan Helgesson, Roberta Ghetti and J\"orgen Ekman \\
   {\small \it School of Technology and Society } \\
   {\small \it Malm\"{o} University, SE-20506 Malm\"{o}, Sweden} }

\date{\today}


\maketitle

\begin{abstract}
From velocity-gated small-angle correlation functions
the emission \\chronology can be deduced for non-identical 
particles, if the emission is independent.
This is not the case for non-identical particles
that originate from two-body decay of fragments.
Experimental results may contain contributions
from both independent emission and two-body decay,
so care is needed in interpreting the velocity-gated 
correlation functions.
It is shown that in some special cases, 
it is still possible to deduce the emission
chronology, even if there is a contribution from
two-body decay.
\end{abstract}

\section{Introduction}
Understanding the different origins of emission and the emission 
time sequence of different particles is a major challenge of 
intermediate energy heavy ion collisions. 
For more than 20 years, 
two-particle intensity interferometry 
has been used as a probe, 
yielding a convoluted space-time information that,
however, is difficult to disentangle
\cite{PLB70_77-Koonin,RMP62_90-Boal,IJMPE6_97-Ardouin,ARNPS49_99-Heinz}. 
This is because of the presence of multiple sources, 
and the competition of non-equilibrium emission 
processes with equilibrium relaxation modes, leading to 
a broad range of origins of the measured particles
\cite{PRC55_97-Larochelle,PRC55_97-Lukasik,PLB491_00-Dore,NPA662_00-Lefort,
      NPA683_01-Lanzano,PLB246_90-Gong,PRC65_02-Verde,PRL75_95-Handzy,
      PRC56_97-Eudes}.
It is the purpose of this paper to discuss some of the
assumptions made when inferring emission chronology
from velocity gated experimental correlation functions.

Particles emitted in nuclear collisions may interact with each other
after the emission. 
This final-state interaction causes the population in
phase-space to be altered, an effect that can be observed
by the small angle two-particle correlation function.
The proximity in space and time of the interacting particles
influences the strength of the final-state interactions,
and hence the size of the correlations seen in the
correlation function $C(q)$, where 
$q\equiv \mu|\mathbf{p}_1/m_1-\mathbf{p}_2/m_2|$ is the relative 
momentum and $\mu = m_1 m_1 / (m_1 + m_2)$ is the reduced mass.
Correlation functions constructed from experimental data
therefore contains information of the space-time characteristics
of the emitting source.
Experimentally, the correlation function
is constructed by dividing the coincidence yield, 
by the yield of non-correlated events, 
normalized to unity at large values of 
relative momentum, where no correlations are expected.
Often, much of the space-time information contained in this  
6-dimensional observable is lost, because 
of implicit experimental integrations over some of the
dimensions in the relative and total momenta. 
When statistics is high enough, 
some of the information can be recovered by applying 
directional cuts \cite{PRC36_87-Pratt} 
and total-momentum or energy gates \cite{RMP62_90-Boal}. 
Furthermore, for non-identical particle correlations, 
it is possible to apply particle-velocity gates and 
infer the emission chronology of the different particles
\cite{PLB373_96-Lednicky,PRL87_01-Ghetti,PRL91_03-Ghetti,
      PRC70_04-Ghetti,NPAsubm-Ghetti,PRCsubm_05-Ghetti}.

Velocity gated correlation functions of unlike particles
is a very powerful method, as it is model independent.
However, there are some critical assumptions made, that
may not always be fulfilled depending 
on the reaction scenario and selection of emitted particles.
For the method to work, the particle velocities should be
obtained in the frame of the emitting source. 
The source velocity may not always be a well defined
quantity, because of implicit  integration over a
limited range of impact parameters. 
In section \ref{Sec:Em-chron-non-id} we investigate 
into some detail, consequences of uncertainties in the source velocity.
Another crucial assumption for extracting emission chronology from
the velocity gated correlation functions,
is that the non-identical particles 
must be emitted {\it independently} by the excited source.
If they instead originate from the two-body decay of a fragment, 
their respective velocity is determined solely by energy and 
momentum conservation, 
and they do not carry any useful information on the emission time. 
In section \ref{Sec:Two-body-decay} we investigate under
which conditions the emission chronology may be
inferred, when particles originate from both independent
emission from a (large) source, and from two-body decay
of fragments.

\section{Space-time characterization}
Normally, particles originating from a specific source 
are emitted at different rates during the time interval of emission, 
leading to a specific time distribution for the source. 
If a two-particle correlation function 
could be constructed from particles emitted from a single source, 
and if the spatial distribution would be known, then
the shape of the correlation function would yield information
on the shape of the time distribution\footnote{
  In Refs. \cite{PRC65_02-Verde,PRC67_03-Verde,PRC64_01-Brown,
                 PLB398_07-Brown,PRC57_98-Brown} 
  the relation between the shape 
  of the correlation function and the shape of the source function
  is discussed.
  If the spatial part of the source is known, 
  the shape relation 
  in Refs. \cite{PRC65_02-Verde,PRC67_03-Verde,PRC64_01-Brown,
                 PLB398_07-Brown,PRC57_98-Brown} 
  is directly applicable to the time distribution.
}.
Normally, the spatial distributions are not known. 
To interpret the experimental results, 
source models, that contain some assumption 
on the spatial and temporal distributions,
are often used.
The shape of these distributions can, to some extent, 
be varied by varying parameters of the model 
The best-fit parameters then represent the average emission point
and average emission time, though it should be remembered
that such average values are model dependent.
In the experimental data the situation is much more complex,
since it is never possible to completely isolate one
source from other sources present during the reaction.
The contribution from several sources leads 
to a complex total time distribution.

The term {\it emission chronology} is usually used to
denote a difference in the {\em average} emission time 
between two particle types.
The difference in the average emission times 
may be small compared with the width of the
emission time distributions.
A difference in the average emission times, 
extracted from experimental data,
may have different origins, 
depending on where the particles, 
that are included in the observables,
are coming from.
Possible origins may be
\begin{itemize}
\item
   A shift of two similar time distributions. 
   This shift could be due to the nuclear interaction.
   For systems with an exotic isospin composition, 
   the symmetry interaction could cause the emission
   time of neutrons and protons to be different.

\item
   A different width of two, otherwise quite similar,  
   time distributions. 
   For an equilibrated source, differences in the Coulomb
   barrier for different particles, could lead to different
   widths of the time distributions.

\item
   Different relative weights of several sources.
   The abundance of emitted neutrons and protons
   from an intermediate velocity source, and
   non-equilibrated residues, could be different
   due to the isospin composition of the systems.

\end{itemize}
When interpreting experimental data, 
it is important to be aware of different possible
origins of different average emission times,
in order to make the correct conclusions.
Different origins can be enhanced or suppressed
by applying different gates and conditions
on the observables.

The emission chronology between two particle types
(e.g.\ neutrons and protons) 
can, under certain conditions, be determined from
like-particle correlation functions (e.g.\ neutron-neutron 
and proton-proton). 
If it is valid to assume that both particle types
are emitted from the same spatial region, 
a model fit to the experimental data, 
will yield an average emission time 
for each particle type.
By comparing these average emission times, 
an emission chronology can be inferred \cite{NPA674_00-Ghetti}. 
The drawback of this method is that 
the results are sensitive to the assumption of emission 
from the same spatial region.
Furthermore, the extracted average emission times
are model dependent, since the average emission
times depend on the shape
of the (spatial and) temporal distributions 
assumed by the specific source model.

\section{Emission chronology from non-identical particle correlations}
\label{Sec:Em-chron-non-id}
Model independent information on the emission
chronology of two particle types (e.g.\ neutrons and protons) can 
be obtained from non-identical-particle correlation functions 
(e.g.\ neutron-proton).
A technique 
was first suggested
for charged particle pairs, based on 
comparison of the velocity difference spectra 
with trajectory calculations 
\cite{NIMA349_94-Gelderloos,PRL75_95-Gelderloos,PRC52_95-Gelderloos}. 
The technique was extended to any kind of interacting, 
non-identical particles,   
by applying energy or velocity gates, 
and proposed for particle pairs such as 
$pd$ and $np$ \cite{PLB373_96-Lednicky}, $p \pi$ \cite{PRL79_97-Voloshin}, 
and $K^+ K^-$ \cite{PLB446_99-Ardouin}. 

\subsection{Particle-velocity-gated correlation functions}
The basic idea of velocity-gated correlation functions of non-identical
particles \cite{PRL87_01-Ghetti,PRL91_03-Ghetti,PRC70_04-Ghetti}, 
is that, if there is an average 
time difference in the emission times of two particles types, 
there will also be a difference in the average distance 
between the particles, 
for particle pairs selected with the condition 
$v_1 > v_2$ as compared to the pairs selected
with the complementary condition  $v_1 < v_2$.
It is obvious that in the class  $v_1 > v_2$
the average velocity of particle 1 will be higher
than in the complementary class  $v_1 < v_2$.
This means that if particle 1 is emitted first,
it will, with the condition  $v_1 > v_2$, 
on average travel a larger distance
before the second particle is emitted,
than with the complementary condition.
In this case the condition $v_1 > v_2$ leads
to on average larger distances 
(and weaker interactions) than the condition $v_1 < v_2$.

The effect can be easily seen if one compares 
the correlation function $C_1$, gated on pairs $v_1 > v_2$, 
with the correlation function $C_2$,
gated on pairs $v_1 < v_2$. 
Assuming that particle 1 is on average emitted first, 
the ratio $C_1/C_2$ will show a dip 
in the region of relative momentum where there is a 
correlation (attractive interaction) 
and a peak where there is an anti-correlation (repulsive interaction).
Furthermore, the ratio $C_1/C_2$ will approach unity 
for both $q \rightarrow 0$ 
(since the velocity difference of the two emitted particles 
is negligible) and $q \rightarrow \infty$
(since modifications of the two-particle phase space density 
arising from final state interactions are negligible). 
A single normalization constant, 
calculated from the non-gated correlation function, 
is utilized for both $C_1$ and $C_2$ \cite{PRL87_01-Ghetti}. 
Note that while the height 
of the non-gated and gated correlation functions 
depends on the normalization, 
and therefore is sensitive to the statistics, 
the ratio of the velocity-gated correlation 
is not sensitive to the normalization, 
since the common normalization constant cancels out.

The exact location of the peak and/or dip in the ratio 
depends on the source and in particular on the origin of
the difference in the average emission times. 
It should also be mentioned that it cannot be ruled out that
the differences observed in velocity-gated correlation functions 
could have a spatial origin.
However, such a correlation would then mean that
there is a correlation between the spatial region
where the particles are emitted and the particle
velocities, {\em and} that this correlation 
is {\em different} for the two particle types 
used in the correlation functions. 
For heavy-ion collisions at intermediate energies, 
such an explanation is clearly more unlike 
than a difference in the average emission times.
%

\subsection{Influence of the source velocity}
\label{Subsec:Influence-source-vel}
The main uncertainty in the method 
of velocity-gated correlation functions 
comes from the uncertainty in the source velocity. 
This uncertainty originates mostly from an
implicit impact parameter averaging in the
experimental data.
The range of selected impact parameters
leads to a distribution of source velocities.
In addition, the measured particles in the pair 
could have been emitted from different sources
with a relatively large difference in source
velocity.
By applying suitable conditions and gates
on the experimental data,
the fraction of such pairs should be low with respect
to particles coming from the desired source.

If the velocity of the assumed source is different
from the real source velocity, the calculated
particle velocities in the assumed source frame will
contain some error, and it may happen that the
magnitude of the two particle velocities 
is interchanged as compared to the real source frame. 
In this section we make an estimate 
of the fraction of pairs that end up
in the wrong gate, depending on the differences
between the real and assumed source velocities.

Assume that the particles 1 and 2 
are emitted with velocities 
$\mathbf{v}_1^{(S)}$ and $\mathbf{v}_2^{(S)}$
from a source S.
If the source has the velocity $\mathbf{v}_s^{(L)}$
in the laboratory system, the particle velocities
in the laboratory system become
\begin{equation}
  \mathbf{v}_i^{(L)} = \mathbf{v}_i^{(S)} + 
                       \mathbf{v}_s^{(L)}, \qquad i=1,2 \ .
\end{equation}
If the velocity $\mathbf{u}_s^{(L)}$ is used to
calculate the particle velocities in the
source system, we get
\begin{eqnarray}
  \mathbf{u}_i^{(S)} 
    & =  &
  \mathbf{v}_i^{(L)} - \mathbf{u}_s^{(L)} \nonumber \\
    & =  &
  \mathbf{v}_i^{(S)} + [ \mathbf{v}_s^{(L)} - \mathbf{u}_s^{(L)} ], 
  \quad i=1,2 \ .
\end{eqnarray}
The condition 
$v_1^{(S)} > v_2^{(S)}$ 
can also be expressed
$0~<~[v_1^{(S)}]^2~-~[v_2^{(S)}]^2$,
and it is straightforward to show that
\begin{eqnarray}
  [u_1^{(S)}]^2 - [ u_2^{(S)} ]^2
    &  =  &
  [v_1^{(S)}]^2 - [ v_2^{(S)} ]^2   \nonumber \\
    &  +  &
    2 [ \mathbf{v}_s^{(L)} - \mathbf{u}_s^{(L)} ] \cdot
    \frac{\mathbf{q}}{\mu} \ .
\label{eq-vel_est}
\end{eqnarray}
The error of using the velocities $u_i$ instead of the 
real velocities $v_i$ can thus be estimated from
Eq.\ (\ref{eq-vel_est}).
Noting that 
$[v_1^{(S)}]^2 - [ v_2^{(S)} ]^2 = 
  [\mathbf{v}_1^{(S)} + \mathbf{v}_2^{(S)}] \cdot \mathbf{q}/\mu$,
the expression shows that if a reasonably good source
selection has been done, so that 
$|\mathbf{v}_s^{(L)} - \mathbf{u}_s^{(L)}|$
is small relative to $|\mathbf{v}_1^{(S)} + \mathbf{v}_2^{(S)}|$, 
the second term in the right hand side of Eq.\ (\ref{eq-vel_est}) 
should be negligible. 
This is especially the case when the relative 
momentum increases, since $|\mathbf{v}_1^{(S)} + \mathbf{v}_2^{(S)}|$ 
increases with $q$ for non-identical particles.
\begin{figure}
\includegraphics[scale=0.55]{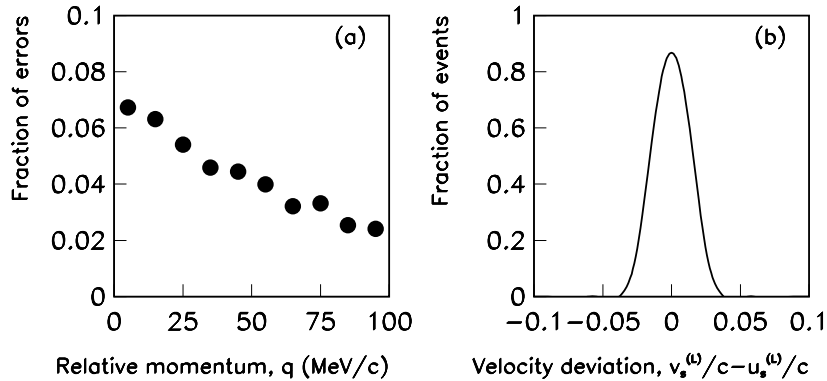}
\caption{\label{fig:vel-err}
         Results from the simulation described in the
         text of subsection \ref{Subsec:Influence-source-vel}. 
         Panel (a) shows the fraction of pairs that
         are attributed to the wrong velocity gate due
         to the uncertainty in the assumed source velocity.
         Panel (b) shows the used distribution of source
         velocities. } 
\end{figure}

As an example we have made a simple simulation for the case
of one proton and one $\alpha$-particle emitted from a source.
We have assumed a Boltzmann distribution of the particle
energies, with temperatures 
$T(p) = 6$ MeV and $T(\alpha)=8 $ MeV.\footnote{ 
  We have assumed that the $\alpha$-particle 
  is emitted somewhat earlier than the proton \cite{NPAsubm-Ghetti}, 
  thus feeling a higher average temperature.
  However, the assumption of different temperatures does not
  influence the conclusions of this section.}
The particles have been assumed to be emitted isotropically.
The difference between the real source velocity and
the assumed one has been taken normally distributed, with
random direction.
The result is presented in Fig.\ \ref{fig:vel-err}a,
where it is seen that the number of pairs attributed
to the wrong velocity gate, is around 5\% for a realistic
estimate of the uncertainty in the source velocity 
(see Fig.\ \ref{fig:vel-err}b).
The error becomes larger if the source velocities
have a larger spread, but even for an unrealistically
large uncertainty in the source velocity, covering
all possible velocities in the reaction, 
the fraction of wrongly attributed pairs is 
less than 25\%.

The example indicates that, for most
cases, the error will be small, of the order
of a few percent.
Nonetheless, this should be explicitly verified in each
application of the method of velocity-gated correlation 
functions.

\section{Two-body Decay from Fragments}
\label{Sec:Two-body-decay}

\subsection{Background}
For those correlation functions 
characterized by final state interactions 
leading to resonances, 
the resonance peaks may a priori have two different origins 
\cite{PRC35_87-Pochodzalla,PRC39_89-DeYoung,PRC41_90-DeYoung}:
\begin{enumerate}
  \item \label{enum-origin-indep}
     From interactions between independently 
     emitted particles from a (large) source\footnote{
       We have here in mind that the two fragments interact
       and thereby change their momenta in such a way that
       the relative momentum region under the resonance peak 
       is populated, but the particle do {\em not} form a fragment
       that then in turn undergo two-body decay. The latter
       process would not be distinguishable from two-body decay
       of fragments formed by other mechanisms in the reaction
       (corresponding to process \ref{enum-origin-frag} in the list).  }. 
    
  \item \label{enum-origin-frag}
     From processes where an unstable fragment formed in the reaction 
     decays into the two measured particles
     (like in e.g.\ $^8$Be $\rightarrow \alpha + \alpha$ 
     or $^6$Li $\rightarrow$ $\alpha$ + $d$).

\end{enumerate}
For the case when a fragment is decaying into two non-identical particles, 
the velocity of the two particles in the fragment rest frame 
is always such that the lighter particle gets a larger velocity 
than the heavier particle. 
This follows from energy and momentum conservation in the decay.

If we could construct a velocity-gated correlation function with
all particles coming from two-body fragment decays and 
with their velocities calculated in the fragment frame, 
we would find all the pairs in the gate 
where the lighter fragment has a higher velocity, 
and none in the complementary gate. 
In subsection \ref{Subsec:Results} we show that
even if the particle velocities are calculated in another
frame, more than 50\% of the events from fragment decay
will still be attributed to the gate 
where the lighter fragment has the higher velocity.

Any deviation from this behavior in experimental data
would be due to particles from other sources 
than two-body fragment decays.
Therefore, when in some cases it is observed that 
the gate where {\it the heaviest} particle has the largest velocity
leads to a stronger correlation or anti-correlation than
the complementary gate,
it can reliably be concluded 
that this behavior is dominated by a mechanism 
different than two-body decay.
In such cases the effect may be attributed
to the interaction of independently emitted particles, 
and the velocity-gated correlation function can be used
to obtain information on the time sequence of the 
independently emitted particles 
(see e.g.\ Ref.\ \cite{NPAsubm-Ghetti}).

\subsection{Kinematics}
\label{Subsec:Kinematics}
Consider a fragment $A_F$ with mass $m_F$ decaying into
two particles $A_1$ and $A_2$, with masses
$m_1$ and $m_2$, 
$A_F~\rightarrow~A_1~+~A_2$.
Momentum conservation in fragment rest frame, 
\begin{equation}
 - \mathbf{p}_1 =  \mathbf{p}_2 \equiv  \mathbf{q} \ ,
\end{equation}  
and energy conservation,
\begin{eqnarray}
 m_F c^2 + E^* 
   & = &        \sqrt{ (m_1 c^2)^2 + (p_1 c)^2 } +
                \sqrt{ (m_2 c^2)^2 + (p_2 c)^2 } \nonumber \\
   & \approx &  m_1 c^2 + \frac{p_1^2}{2 m_1} +
                m_2 c^2 + \frac{p_2^2}{2 m_2} \ ,
\end{eqnarray}  
lead to
\begin{eqnarray}
  (q c)^2 
    & = &       \frac{(m_F c^2 + E^*)^2}{4} + 
                \frac{(m_1^2 c^4 - m_2^2 c^4)^2}
                     {4(m_F c^2 + E^*)^2} -
                \frac{m_1^2 c^4 + m_2^2 c^4}{2}  \nonumber \\
    & \approx & \frac{2 m_1 c^2 m_2 c^2 (m_F c^2 + E^*)}
                     {m_1 c^2 + m_2 c^2} - 2 m_1 c^2 m_2 c^2 \ ,
\end{eqnarray}  
where $q$ is the relative momentum of the two particles,
and $E^*$ is the excitation energy of the decaying fragment.
In the rest frame of the decaying fragment, the velocities
of the two emitted particles become
\begin{eqnarray}
  \mathbf{v}_1^{(F)} & = & -\mathbf{q} / m_1   \\
  \mathbf{v}_2^{(F)} & = &  \mathbf{q} / m_2   \ ,
\end{eqnarray}
and the velocity of the lighter particle 
will always be larger than the velocity of the heavier particle.

If the velocity of the fragment in the laboratory system 
is $\mathbf{v}_F^{(L)}$, then the particle velocities in the 
laboratory system become
\begin{eqnarray}
  \mathbf{v}_1^{(L)} & = & \mathbf{v}_F^{(L)} + \mathbf{v}_1^{(F)} 
                       =   \mathbf{v}_F^{(L)} - \mathbf{q} / m_1   \\
  \mathbf{v}_2^{(L)} & = & \mathbf{v}_F^{(L)} + \mathbf{v}_2^{(F)} 
                       =   \mathbf{v}_F^{(L)} + \mathbf{q} / m_2   \ .
\end{eqnarray}
In applications of velocity-gated correlation functions,
the particle velocities are calculated in some source system
(intermediate velocity, target or projectile residue source) 
with assumed velocity $v_s^{(L)}$. 
The particle velocities in the source system become
\begin{eqnarray}
  \mathbf{v}_1^{(S)} & = & \mathbf{v}_1^{(L)} - \mathbf{v}_S^{(L)} 
                       =   \mathbf{v}_F^{(L)} - \mathbf{q} / m_1
                       -   \mathbf{v}_S^{(L)} \\
  \mathbf{v}_2^{(S)} & = & \mathbf{v}_2^{(L)} - \mathbf{v}_S^{(L)} 
                       =   \mathbf{v}_F^{(L)} + \mathbf{q} / m_2
                       -   \mathbf{v}_S^{(L)} \ .
\end{eqnarray}

The condition of the velocity gates, $v_1 > v_2$ 
($v_1 < v_2$), 
can also be written
$v_1^2 - v_2^2 > 0$ ($v_2^1 - v_2^2 < 0$).
Using the particle velocities in the source system,
we get 
\begin{equation}
  (v_1^{(S)})^2 - (v_2^{(S)})^2 
     = 
        \frac{q}{\mu}\left[ \frac{m_2-m_1}{m_1 m_2} \, q 
     -   2 \, \mathbf{e}_q \cdot \mathbf{v}_F^{(S)} \right] \ ,
\label{eq_vel-cond}
\end{equation}
where $\mathbf{e}_q$ is a unit vector directed along $q$,
and $\mathbf{v}_F^{(S)} = \mathbf{v}_F^{(L)} - \mathbf{v}_S^{(L)}$
is the fragment velocity in the source frame.
In a given decay, $q$, $m_1$ and $m_2$ are determined by energy and
momentum conservation, while the direction of $\mathbf{q}$ can be
considered to be isotropic.
If we denote the angle between $\mathbf{q}$ and $\mathbf{v}_F^{(S)}$
by $\beta$, the condition in Eq.\ (\ref{eq_vel-cond}) can be written
\begin{equation}
  (v_1^{(S)})^2 - (v_2^{(S)})^2 
    =  \frac{q}{\mu}\left[ \frac{m_2-m_1}{m_1 m_2} \, q 
        - 2 \, v_F^{(S)} \cos(\beta) \right].
\label{eq:vel-cond2}
\end{equation}

\subsection{Results}
\label{Subsec:Results}
From Eq.\ (\ref{eq:vel-cond2}) it is clear that in the source
system, the light particle does not always have a larger 
velocity than the heavier.
The condition depends on the source velocity, and on the
angle between the relative momentum and the source velocity.
\begin{figure}[t]
\includegraphics[scale=0.35]{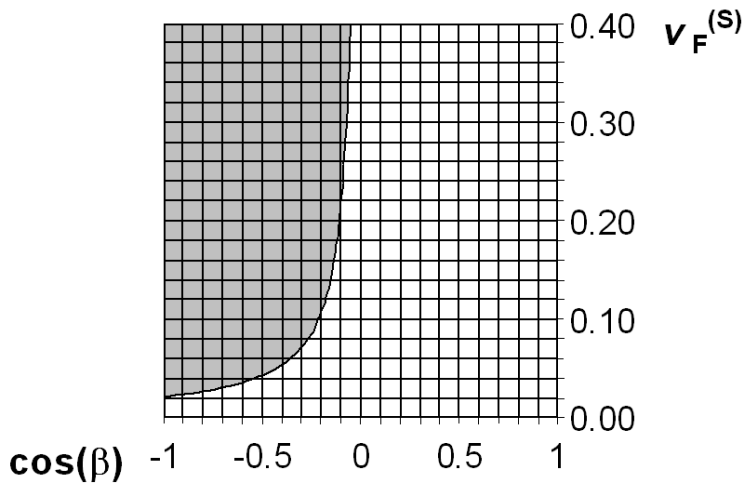}
\caption{\label{fig:vcond1} 
         A graphical representation of the velocity condition 
         in Eq.\ (\ref{eq:vel-cond2}) for the decay 
         $^5$Li $\rightarrow p$ + $\alpha$. 
         The gray area represents the region 
         where $v^2_p - v^2_{\alpha} < 0$.}
\end{figure}

In Fig.\ \ref{fig:vcond1} we present the regions where
the lighter or heavier particle has the largest velocity.
When the fragment velocity (in the source system) is small,
the velocity condition in the fragment frame still holds 
(i.e.\ the lighter particle has the higher velocity).
But, as the difference between the fragment velocity and
the source velocity becomes larger, a larger fraction
of events will have the velocity condition interchanged.
However, it is important to note that the number of events
for which the original velocity condition holds is 
always more than 50\%
assuming that $\mathbf{q}$ is isotropically distributed.
This means that if all particle pairs would
come from two-body fragment decays, then the correlations
in the gate where the lighter particle has the higher 
velocity (in whatever frame) will always be stronger 
than in the complementary gate.
Thus if the contrary behavior is observed, the 
dominating origin of particles must be other than 
two-body decay.

From Eq.\ \ref{eq:vel-cond2} it is clear that the
regions in the source system (as in Fig.\ \ref{fig:vcond1})
where lighter or heavier particle 
has the largest velocity, depends on the
relative momentum, $q$, in the decay.
For two-body decay that yields large
values of $q$, the region where the
lighter particle has the largest velocity
will be larger than for decays yielding smaller
values of $q$.
This dependence is the main reason for different
results of the subsections 
\ref{Subsubsec:PLS-emission} 
and
\ref{Subsubsec:IS-emission}
below.

In experimental data, the situation is
somewhat more complicated, because of
limited angular coverage.
The angular coverage may select certain regions
in the $\cos(\beta)$~-~$v_F^{(S)}$ plane, that
then could change the above conclusions.
It is not possible to find a simple expression
for the velocity condition including the
angular coverage of a given experiment.
Instead this has to be investigated numerically
for each setup whenever velocity gates are used.
In the next two sections we presents the
results for two such investigations, and show
how they could be implemented.

\subsubsection{Application to emission from projectile residue}
\label{Subsubsec:PLS-emission}
\begin{figure}[t]
\includegraphics[scale=0.45]{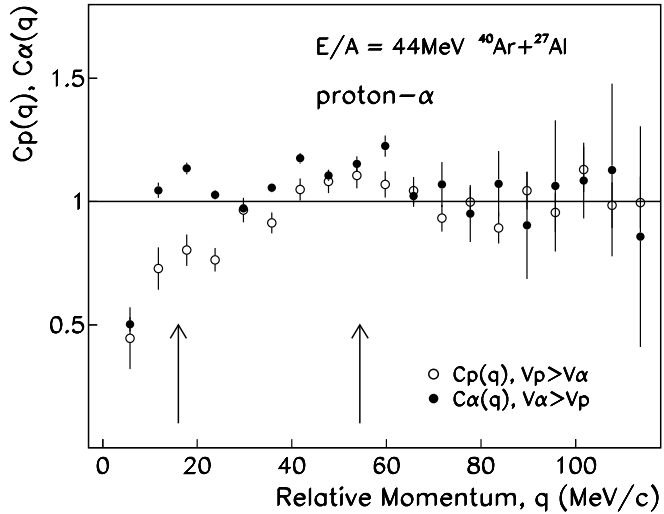}
\caption{\label{fig:Cpa} 
         The experimental velocity-gated $p \, \alpha$
         correlation function of Ref.\ \cite{NPAsubm-Ghetti}. 
         The arrows in the figure indicates the positions
         of different fragment decays (see Ref.\ \cite{NPAsubm-Ghetti}). 
         } 
\end{figure}

In Ref.\ \cite{NPAsubm-Ghetti} velocity-gated small angle 
two-particle
correlation functions were used for 
protons, deuterons, tritons and $\alpha$-particles, from the
$E/A$ = 44 and 77 MeV $^{40}$Ar + $^{27}$Al collisions,
at very forward angles, with the aim of studying emission
from the projectile-like residue (PLS).
Fig.\ \ref{fig:Cpa} shows  
velocity-gated $p \, \alpha$-correlation functions
for the $E/A$ = 44 MeV reaction.
One can note that in the region $q \sim 54$ MeV$/c$,
corresponding to the two-body decay of $^5$Li,
the gate $v_{\alpha} > v_p$ shows larger correlations
than the complementary gate $v_{\alpha} < v_p$.
One possible deduction from this observation 
(and the discussion from the previous section) 
would be that these particle must have a different origin
than decay from $^5$Li.
However, before such a conclusion can be drawn, 
it must be ruled out that the very
specific angular coverage of $0.7^{\circ} < \theta < 7^{\circ}$
in Ref.\ \cite{NPAsubm-Ghetti} does not impose such
constraints in the $\cos(\beta)$~-~$v_F^{(S)}$ plane
of Fig.\ \ref{fig:vcond1}, that the velocity
condition is interchanged.

To investigate this issue, we have performed simple
numerical simulations, where we have assumed
that the proton and the $\alpha$-particle come
from decay of $^5$Li, and that the $^5$Li has
been emitted from different sources.
We require that the proton and the $\alpha$-particle 
should come within the angular range of the
detectors ($0.7^{\circ} < \theta < 7^{\circ}$),
as well as be above the energy threshold
of 35 A MeV, imposed in Ref.\ \cite{NPAsubm-Ghetti}.
The results are summarized in Table \ref{tab:pa}.
The first row shows the results when we assume that the
$^5$Li has been emitted from a target-like residue source (TLS)
of temperature 7 MeV. We deduce the kinetic energy 
of the $^5$Li from a Boltzmann distribution, 
assume that the emission is isotropic in the TLS frame,
and boost the $^5$Li with the TLS velocity in the 
laboratory frame. The TLS velocity is taken 
normally distributed around the mean value of 0.02 $c$ with
a standard deviation of 0.01 $c$.
The decay of $^5$Li is assumed to be isotropic,
while the magnitude of the proton and $\alpha$ velocities
in the $^5$Li frame is determined 
from the energy and momentum conservation in the decay.
Finally the proton and $\alpha$ velocities are
calculated in the frame of the projectile-like residue source 
(PLS), which is assumed to have the velocity 0.27 $c$ along the
beam direction.
We have used 200 000 events in the simulations.
The second column in Table \ref{tab:pa} shows in how many
of {\em all} events, irrespectively of detection, 
that the proton has a larger velocity in the PLS-frame,
than the $\alpha$-particle.
The third column shows in how many of the events that
both particles fall within the angular range 
of the detectors, and above the imposed energy threshold.
The fourth column shows in how many of the events 
in column three, that the proton has a higher velocity
in the PLS-frame, than the $\alpha$-particle.
For the intermediate velocity source (IS) and the 
PLS we have assumed temperatures of 20 and 7 MeV, respectively,
and source velocities normally distributed around the
mean values 0.15 $c$ and 0.27 $c$, respectively, with
standard deviations of 0.02 and 0.01.
\begin{table}[t]
\begin{tabular}{|c|c|c|c|}
\hline
\hline
Fragment  &  All                            & \multicolumn{2}{c|}{$0.7 < \theta_{lab} < 7.0^\circ$} \\
source    &  pairs                          & \multicolumn{2}{c|}{$E_k/A > 35$ MeV} \\
          &  $v_p^{(PLS)} > v_{\alpha}^{(PLS)}$    & Detected                                              & $v_p^{(PLS)} > v_{\alpha}^{(PLS)}$ \\
\hline
TLS       & 54.3\% & 0.00\% & \\
IS        & 58.6\% & 0.03\% & 84.7\% \\
PLS       & 76.9\% & 2.27\% & 82.8\% \\
\hline
\hline
\end{tabular}
\caption{\label{tab:pa}The fraction of $p \alpha$ pairs
emitted in the two-body decay of $^5$Li which have
$v_p^{(PLS)} > v_{\alpha}^{(PLS)}$.
The last column gives the fraction of ``detected'' pairs.}
\end{table}

As seen in Table \ref{tab:pa}, the experimental filter
in this case strengthens the velocity condition
in the $^5$Li frame, as compared to the unfiltered
result (column two).
This means that if the protons and $\alpha$-particles 
in the region $q \sim 54$ MeV$/c$,
would come entirely from decay of $^5$Li,
more than 80\% of the pairs would come
in the gate $v_p^{(PLS)} > v_{\alpha}^{(PLS)}$
and this correlation function would be much
higher than the complementary gate. 
Since this behavior is not observed, it should be safe 
to conclude that the protons and $\alpha$-particles 
contributing to the correlations in Fig.\ \ref{fig:Cpa}
(in the region $q \sim 54$ MeV$/c$)
must have another origin than decay of $^5$Li.
In Ref.\ \cite{NPAsubm-Ghetti}, it is assumed that these
particles are emitted independently from the
PLS source, and that the correlations originate 
from final state interaction between particles
emitted close in space and time.

\subsubsection{Application to intermediate velocity source emission}
\label{Subsubsec:IS-emission}

\begin{figure}[t]
\includegraphics[scale=0.45]{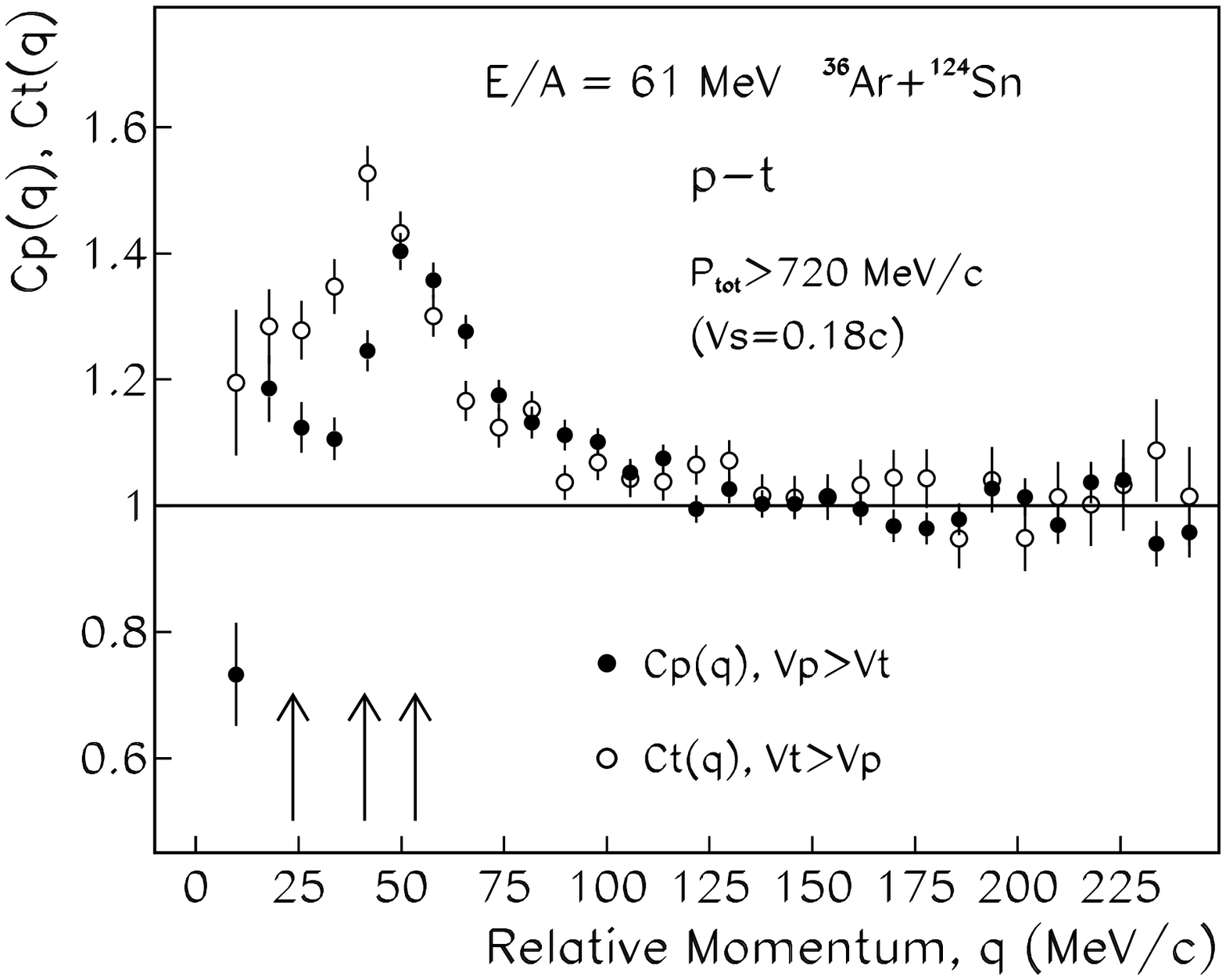}
\caption{\label{fig:Cpt} 
         The experimental velocity-gated $p \, t$ 
         correlation of Ref.\ \cite{PRCsubm_05-Ghetti}.
         The arrows in the figure indicates the positions
         of different fragment decays (see Ref.\ \cite{PRCsubm_05-Ghetti}).
         }
\end{figure}

In this section we present numerical simulations
similar to those presented in section \ref{Subsubsec:PLS-emission},
but now for a different reaction and experimental filter.
In Ref.\ \cite{PRCsubm_05-Ghetti}  
velocity-gated small angle two-particle
correlation functions were used to deduce the 
emission chronology of protons, deuterons, and tritons, from the
$E/A$ = 61 MeV $^{36}$Ar + $^{112,124}$Sn collisions. 
Figure \ref{fig:Cpt} illustrates  
velocity-gated $p \, t$-correlation functions
for the $^{36}$Ar + $^{124}$Sn reaction.
In this case the angular coverage is
$30^{\circ} < \theta < 114^{\circ}$,
and the source of particle emission is assumed
to be the intermediate velocity source (IS). 
In particular, we have used the parameters summarized
in Table \ref{tab:pt-sim}.

\begin{table}[b]
\begin{tabular}{|c|c|c|c|}
\hline
\hline
Fragment  &  T    & $<v>/c$ & $\sigma_v/c$ \\ 
source    & (MeV) &         &              \\
\hline
TLS       &  7    &  0.01   &   0.01       \\
IS        & 20    &  0.18   &   0.02       \\
PLS       & 7     & 0.31    &   0.01       \\
\hline
\hline
\end{tabular}
\caption{\label{tab:pt-sim}Parameters used in the simulation 
of section \ref{Subsubsec:IS-emission}.}
\end{table}
The results of the simulations are presented in Table \ref{tab:pt}.
We see that also in this case the 
velocity condition from the $^4$He frame
holds, even though for a substantial part
of the pairs ($\sim$ 43\%) 
the triton gets a higher velocity than the proton
in the IS frame.
This means that if the protons and tritons
in the region $q \sim 24$ MeV$/c$,
would come entirely from decay of $^4$He,
then more than 56\% of the pairs would come
in the gate $v_p^{(IS)} > v_t^{(IS)}$
and this correlation function would be 
higher than the complementary gate. 
Since this is not observed it should 
also in this case be safe 
to conclude that the protons and tritons
contributing to the correlations of Fig.\ \ref{fig:Cpt}
(in the region $q \sim 24$ MeV$/c$)
must have a different origin than decay of $^4$He.
In Ref.\ \cite{PRCsubm_05-Ghetti} it is assumed that these
particles are emitted independently from the
IS source, and that the correlations originate 
from final state interaction between particles
emitted close in space and time.

\section{Summary}
Two-particle intensity interferometry is an important tool to access information 
on the space-time characteristics of particle emitting sources at intermediate 
energy heavy ion collisions. In particular, when pairs of non-identical particles 
are detected in coincidence, particle-velocity-gated correlation functions can be used 
to establish the emission time sequence, in a model-independent way. 
Since the particle-velocities have to be calculated in the frame of the 
emitting source, the main uncertainty in the method 
comes from the uncertainty in the source velocity, mostly originating from 
impact parameter averaging implicitly performed in experiments.
We have demonstrated how the error due to uncertainties
in the source velocity can be estimated by numerical simulations.
A presented example of such an investigation, makes it plausible
that for most cases, the error will be small, 
of the order of a few percent.  
\begin{table}[t]
\begin{tabular}{|c|c|c|c|}
\hline
\hline
Fragment  &  All                            & \multicolumn{2}{c|}{$30 < \theta_{lab} < 114^\circ$} \\
source    &  pairs                          & \multicolumn{2}{c|}{$E_k > 10$ MeV} \\
          &  $v_p^{(PLS)} > v_a^{(PLS)}$    & Detected                                              & $v_p^{(PLS)} > v_a^{(PLS)}$ \\
\hline
TLS       & 52.5\% & 0.16\% & 56.4\% \\
IS        & 56.5\% & 14.97\% & 56.9\% \\
PLS       & 53.2\% & 0.52\% & 58.9\% \\
\hline
\hline
\end{tabular}
\caption{\label{tab:pt}The fraction of $p \, t$ pairs
emitted in the two-body decay of $^4$He which have
$v_p^{(PLS)} > v_t^{(PLS)}$.
The last column gives the fraction of ``detected'' pairs.}
\end{table}

A more serious problem may be posed by the fact that
for the particle-velocity-gated correlation function 
method to work, the particles have to be 
emitted independently by the source. 
Clearly, this is not always the case, 
as particle pairs may originate from two-body decay of excited fragments. 
We have demonstrated that the kinematical signature of two-body 
decay is strong in the particle-velocity-gated correlation functions. 
This leads to an expected behavior, 
namely that the correlations in the gate where the lighter particle has 
the higher velocity are stronger than in the complementary gate.
This ensures that when the contrary behavior is observed, the 
dominating origin of particles must be other than 
two-body decay. As an example, we have shown two experimental 
cases where it has been possible to deduce the emission chronology, 
even if there is a contribution from two-body decay.




\end{document}